\documentclass[aps,prl,twocolumn,showpacs]{revtex4}

\usepackage{amssymb,amsmath,graphicx}

\begin{document}

\title{The inhomogeneous evolution of subgraphs and cycles in complex 
networks}

\author{$^1$Alexei V\'azquez, $^{1,2}$J. G. Oliveira, and 
$^1$Albert-L\'aszl\'o Barab\'asi}

\affiliation{$^1$Department of Physics and Center for Complex Network
Research, University of Notre Dame, IN 46556, USA}

\affiliation{$^2$Departamento de F\'isica, Universidade de Aveiro, Campus
Universit\'ario de Santiago, 3810-193 Aveiro, Portugal}

\date{\today}

\begin{abstract}

Subgraphs and cycles are often used to characterize the local properties
of complex networks. Here we show that the subgraph structure of real
networks is highly time dependent: as the network grows, the density of
some subgraphs remains unchanged, while the density of others increase at
a rate that is determined by the network's degree distribution and
clustering properties. This inhomogeneous evolution process, supported by
direct measurements on several real networks, leads to systematic shifts
in the overall subgraph spectrum and to an inevitable overrepresentation
of some subgraphs and cycles.

\end{abstract}


\maketitle

\bibliographystyle{apsrev}


Subgraphs, representing a subset of connected vertices in a graph,
provide important information about the structure of many real networks.
For example, in cellular regulatory networks feed-forward loops play a
key role in processing regulatory information \cite{alon02},
while in protein interaction networks highly connected subgraphs
represent evolutionary conserved groups of proteins \cite{wuchty03}. In a
similar vain, cycles, a special class of subgraphs, offer evidence for
autonomous behavior in ecosystems \cite{ulanowicz83}, cyclical exchanges
give stability to social structures \cite{bearman97}, and cycles
contribute to reader orientation in hypertext \cite{bernstein00}.
Finally, understanding the nature and frequency of cycles is important
for uncovering the equilibrium properties of various network models
\cite{marinari04}.

Motivated by these practical and theoretical questions, recently a series
of statistical tools have been introduced to evaluate the abundance of
subgraphs \cite{alon02,wuchty03,vazquez04} and cycles
\cite{bianconi03a,bianconi04,rozenfeld04,sergi04}, offering a better 
description
of a network's local organization. Yet, most of these methods were
designed to capture the subgraph structure of a specific snapshot of a
network, characterizing static graphs. Most real networks, however, are
the result of a growth process, and continue to evolve in time
\cite{ab01a}. While growth often leaves some of the network's global
features unchanged, it does alter its local, subgraph based structure,
potentially modifying everything from subgraph densities to cycle
abundance. Yet, the currently available statistical methods cannot
anticipate or describe such potential changes.

In this paper we show that during growth the subgraph structure of
complex networks undergoes a systematic reorganization. We find that the
evolution of the relative subgraph and cycle abundance can be predicted
from the degree distribution $P(k)$ and the degree dependent average
clustering coefficient $C(k)$. The results indicate that the subgraph
composition of complex networks changes in a very inhomogeneous manner:
while the density of many subgraphs is independent of the network size,
they coexist with a class of subgraphs whose density increases at a
subgraph dependent rate as the network expands. Therefore in the
thermodynamic limit a few subgraphs will be highly overrepresented
\cite{alon02}, a prediction that is supported by direct
measurements on a number of real networks for which time resolved network
topologies are available. This finding questions our ability to
characterize networks based on the subgraph abundance obtained from a
single topological snapshot. We show that a combined understanding of
network evolution and subgraph abundance offers a more complete picture.

\begin{figure}
\centerline{\includegraphics[width=2in]{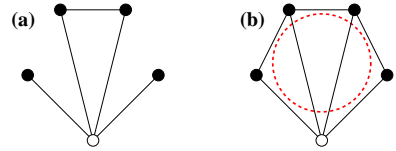}}

\caption{Examples of subgraphs and cycles with a central vertex. The
subgraph shown in (a) has $n=5$ vertices and $n-1+t=5$ edges, where $t=1$
represents the number of edges connecting the neighbors of the central
vertex (empty circle) together.  In (b) we show a subgraph with $t=3$
edges among the neighbors, such that the central vertex and its neighbors
form a cycle of length $h=5$, highlighted by the dotted circle}

\label{fig1}
\end{figure}

{\it Subgraphs}: We consider subgraphs with $n$ vertices and $n-1+t$
edges, whose central vertex has links to $n-1$ neighbors, which in turn
have $t$ links among themselves (Fig. \ref{fig1}a).  The total number of
$n$-node subgraphs that can pass by a node with degree $k$ is
$\binom{k}{n-1}$. Each of these $n$-node subgraphs can have at most
$n_{\textrm{p}}=(n-1)(n-2)/2$ edges between the $n-1$ neighbors of the
central node. The probability that there is an edge between two neighbors
of a degree $k$ vertex is given by the clustering coefficient $C(k)$.
Therefore, the probability to obtain $t$ connected pairs and
$n_{\textrm{p}}-t$ disconnected pairs is given by the binomial
distribution of $n_p$ trials with probability $C(k)$. The expected number
of ($n,t$) subgraphs in the network is obtained after averaging over the
degree distribution, resulting in

\begin{equation}
N_{nt}=g_{nt}N\sum_{k=1}^{k_{\max}}P(k)\binom{k}{n-1}
\binom{n_{\textrm{p}}}{t}C(k)^t[1-C(k)]^{n_{\textrm{p}}-t}\ ,  
\label{Nnt}
\end{equation}

\noindent where $k_{\max}$ is the maximum degree and the geometric factor
$g_{nt}$ takes into account that the same subgraph can have more than one
central vertex. For instance, a triangle will be counted three times since
each vertex is connected to the others, therefore $g_{31}=1/3$. For
networks where $P(k)\sim k^{-\gamma}$ and $C(k)\sim k^{-\alpha}$, where
$\gamma$ and $\alpha$ are the degree distribution and clustering hierarchy
exponents, in the thermodynamic limit $k_{max}\rightarrow\infty$ Eq.
(\ref{Nnt}) predicts the existence of two subgraph classes
\cite{vazquez04}

\begin{equation}
\frac{N_{nt}}{N}\sim
\left\{
\begin{array}{lll}
C_0^t k_{\max}^{n-\gamma-\alpha t}\ , & n-\gamma-\alpha t>0\ ,
& \mbox{Type I} \ ,\\
C_0^t \ , & n-\gamma-\alpha t<0\ ,
& \mbox{Type II} \ .
\label{NI}
\end{array}
\right.
\end{equation}

\begin{table}
\begin{tabular}{|l|l|l|l|l|l|l|l|}
\hline
Network & $\gamma$ & $\alpha$ & $\delta$ & $\theta$ & $\zeta_3$ & 
$\zeta_5$ & $\zeta_5$\\
\hline
Co-authorship & 2.4 & 0.0 & 0.6 & 0.00 & 0.6 & 1.6 & 2.6\\
Internet & 2.2 & 0.75 & 1.0 & 0.20 & 0.3 & 0.7 & 1.2\\
Language & 2.7 & 1.0 & 0.40 & 0.68 & 0.7 & 1.4 & 2.0\\
Model & 2.6 & 1 & 0.63 & 0 & 0 & 0 & 0\\
\hline
\end{tabular}

\caption{Characteristic exponents of the investigated real networks and
the deterministic model. The exponents are defined through the scaling of
the degree distribution $P(k)\sim k^{-\gamma}$, the clustering
coefficient $C(k)=C_0k^{-\alpha}$, with $C_0\sim N^\theta$, the largest
degree $k_{max}\sim N^\delta$, and the number of $h$-cycles $N_h/N\sim
N^{\zeta_h}$.}

\label{tab1}
\end{table}

\noindent Therefore, for the Type I subgraphs the $N_{nt}/N$ density
increases with increasing network size, and $N_{nt}/N$ is independent of
$N$ for Type II subgraphs. In the following we provide direct evidence for the two subgraph
types in several real networks for which varying network sizes are
available: co-authorship network of mathematical publications
\cite{bjnr01a}, the autonomous system representation of the Internet
\cite{pvv01,vpv02a}, and the semantic web of English synonyms
\cite{yook}. In each of these networks the maximum degree increases
as $k_{max}\sim N^\delta$. We estimated $\delta$ from the scaling of the
degree distribution moments with the graph size, $\left<k^n\right>\sim
N^{\delta(n+1-\gamma)}$, with $n=2,3,4$. Furthermore, we find that $C_0$
from $C(k)=C_0k^{-\alpha}$ also depends on the network size as $C_0\sim
N^\theta$, where $\theta$ can be estimated using
$C_0=\sum_{k\geq2}C(k)/\sum_{k\geq2}k^{-\alpha}$, giving a better
estimate than a direct fit of $C(k)$. The exponents characterizing each
network are summarized in Table. \ref{tab1}.

\begin{figure}
\centerline{\includegraphics[width=2.9in]{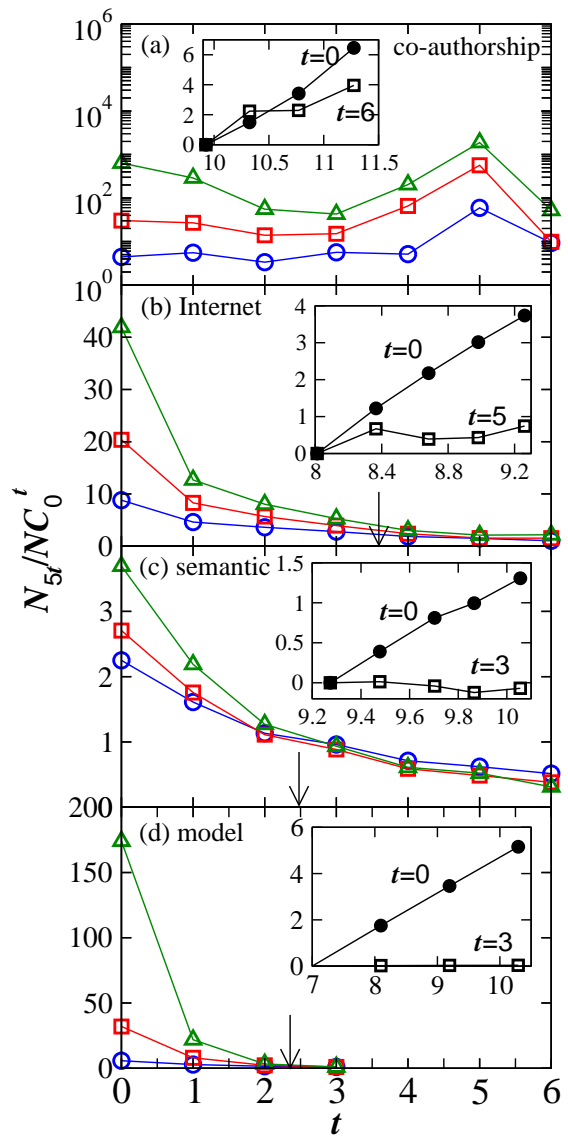}}

\caption{Number of ($n=5$,$t$) subgraphs for the co-authorship (a),
Internet (b), semantic (c) networks and the deterministic model (d) as a
function of $t$. Different symbols correspond to different snapshots of
the networks evolution, from early stage (circles) to intermediate
(squares)  and current ({\it i.e.} largest) (triangles). $N_{nt}$ depends
strongly on $t$ (spanning several orders of magnitude) making difficult to
observe the $N$ dependence. Thus we normalized all the quantities
($N_{5t}$, $C_0$ and $N$)  to the first year available. The arrows
correspond to the phase boundary $5-\gamma-\alpha t=0$, with Type I and II
subgraphs to the left and right of the arrow, respectively. In the insets
show the system size dependence we plot $\log N_{5t}$ vs $\log N$ for
different values of $t$.}

\label{fig2}
\end{figure}

In Fig. \ref{fig2} we show the density of all five vertex subgraphs
($n=5$)  as a function of $t$.  For the Internet and Language networks
$C_0$ increases with $N$, therefore the subgraph's density increases with
the network size for all subgraphs. This consequence of the
non-stationarity of the clustering coefficient is subtrated by
normalizing $N_{nt}$ by $C_0^t$. For the co-authorship graph with
$\alpha=0$ (Table \ref{tab1}), only Type I subgraphs are observed, as
predicted by (\ref{NI}). In contrast, for the Internet and semantic
networks $\alpha>0$, therefore the overrepresented Type I phase is
expected to end approximately at the phase boundary predicted by
(\ref{NI}). Indeed, left to the arrow denoting the $n-\gamma-\alpha t$
phase boundary we continue to observe a systematic increase in
$N_{5t}/NC_0^t$, as expected for Type I subgraphs. In contrast, beyond the
phase boundary the subgraph densities obtained for different network sizes
are independent of $N$, collapsing into a single curve.

We compared our predictions with direct counts in a growing deterministic
network model \cite{dgm02a} as well, characterized by a degree exponent
$\gamma=1+\ln3/\ln2\approx2.6$ and a degree dependent clustering
coefficient $C(k)=C_0k^{-\alpha}$, with $C_0=2$ and $\alpha=1$. In Fig.
\ref{fig2}d we show the number of ($n=5$,$t$)  subgraphs for different
values of $t$ and graph sizes. The arrow indicating the predicted phase
transition point $n-\gamma-\alpha t =0$ clearly separates the Type I from
the Type II subgraphs, a numerical finding that is supported by exact
calculations as well. Note that only one Type II $n=5$ subgraph is
present in the deterministic network, due to its particular evolution
rule.

{\it Cycles}: The formalism developed above can be generalized to predict
cycle abundance as well. Consider the set of centrally connected
cycles shown in Fig. \ref{fig1}b.  If the central vertex has degree $k$,
we can form $\binom{k}{h-1}$ different groups of $h$ vertices, $h-1$
selected from its $k$ neighbors and the central vertex. Each ordering of
the $h-1$ selected neighbors corresponds to a different cycle, therefore
we multiply with half of the number of their permutations $(h-1)!$
(assuming that 123 is the same as 321). Finally, to obtain the number of
$h$-cycles we multiply the result with the probability of having $h-2$
edges between consecutive neighbors, $C(k)^{h-2}$, and sum over the
degree distribution $P(k)$, finding

\begin{equation}
\frac{N_h}{N} = g_h \sum_{k=h-1}^{k_{max}} P(k) \frac{(h-1)!}{2} 
\binom{k}{h-1} C(k)^{h-2}\ ,
\label{nhb}
\end{equation}

\noindent where $g_h$ is again a geometric factor correcting multiple
counting of the same cycle. Note that (\ref{nhb}) represents a lower
bound for the total number of $h$-cycles, which also include cycles
without a central vertex. Depending on the values of $h$, $\gamma$ and
$\alpha$ the sum in (\ref{nhb}) may converge or diverge in the limit
$k_{max}\rightarrow\infty$. When it converges, the density of $h$-cycles
is independent of $N$ (Type II), otherwise it grows with $N$ (Type I).
Since in preferential attachment models without clustering the density of
$h$-cycles decreases with increasing $N$ \cite{bianconi03}, we conclude
that clustering is the essential feature that gives rise to the observed
high $h$-cycle number in such real networks like the Internet
\cite{bianconi03a}. To further characterize the cycle spectrum, we need
distinguish two different cases, $0<\alpha<1$ and $\alpha\geq1$.

\begin{figure}
\centerline{\includegraphics[width=3in]{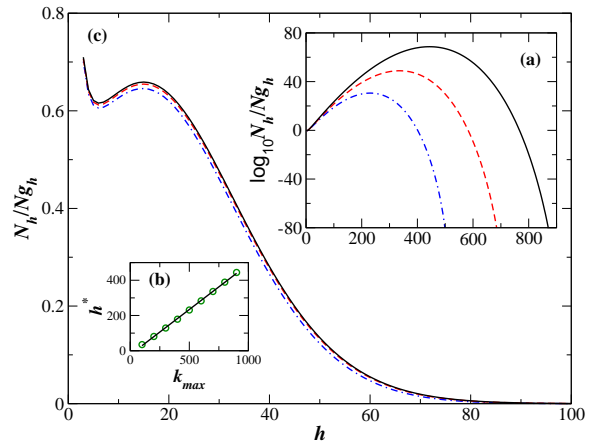}}

\caption{Number of $h$-cycles as computed from (\ref{nhb}), using
$\gamma=2.5$, (a) $C_0=1$ and $\alpha=0.9$, (b) $C_0=2$ and $\alpha=1.1$,
and $k_{max}=500$ (dashed-dotted), 700 (dashed) and 900 (solid). (c) $h$
value at which $N_h$ has a maximum as a function of $k_{max}$.}

\label{fig3}
\end{figure}

$0<\alpha<1$: In the $k_{max}\rightarrow\infty$ limit the 
cycle density follows

\begin{equation}
\frac{N_h}{N}\sim \left\{
\begin{array}{ll}
C_0^{h-2}\ , & h<h_c\ ,\\
C_0^{h-2} k_{max}^{(1-\alpha)(h-h_c)}\ , & h>h_c\ ,
\end{array}
\right.
\label{nha1}
\end{equation}

\noindent where $h_c=(\gamma-2\alpha) / (1-\alpha)$. Therefore, large
cycles ($h>h_c$) are abundant, their density growing with the network size
$N$. As $\alpha\rightarrow1$ the threshold $h_c\rightarrow\infty$,
therefore the range of $h$ for which the density is size independent
expands significantly.

Direct calculations using (\ref{nhb}) show that $N_h$ exhibits a maximum
at some intermediate value of $h$ (see Fig. \ref{fig3}a, already reported
for the deterministic model \cite{rozenfeld04}. The maximum represents a
finite size effect, as the characteristic cycle length $h^*$,
corresponding to the maximum of $N_h$, scales as $h^* \sim k_{max}$ (Fig.
\ref{fig3}b). Yet, next we show that this behavior is not generic, but
depends on the value of $\alpha$.

$\alpha\geq1$: For all $\gamma>2$ only Type II subgraphs are expected
($N_h/N\sim C_0^{h-2})$, as suggested by the divergence of $h_c$ in the
$\alpha\rightarrow1$ limit.  If $C_0>1$ the number of $h$-cycles continues
to exhibit a maximum and the characteristic cycle length $h^*$ scales as
$h^*\sim k_{max}$. If $C_0<1$, however, the number of $h$-cycles decrease
with $h$, although a small local minima is seen for small cycles. More
important, in this case $N_h/N$ is independent of the network size (see
Fig. \ref{fig3}c), in contrast with the size dependence observed earlier
(Fig.  \ref{fig3}a and \cite{rozenfeld04}). Thus, for networks with
$\alpha>1$ or $\alpha=1$ and $C_0<1$ the cycle spectrum is stationary,
independent of the stage of the growth process in which we inspect the
network.

\begin{figure}
\centerline{\includegraphics[width=3in]{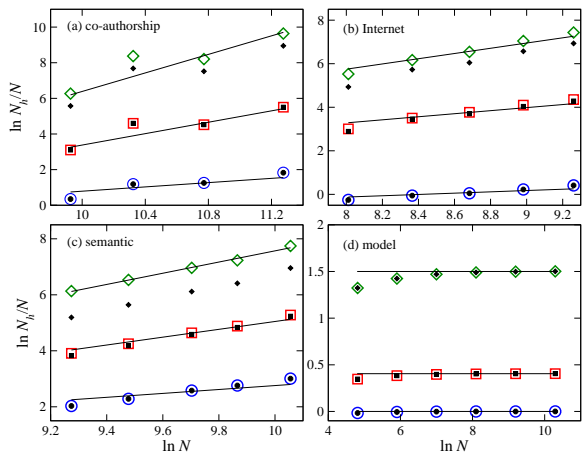}}

\caption{Density of all (open symbols) and centrally connected (filled
symbols) cycles with $h=3$ (circles), 4 (squares) and 5 (diamond)  
cycles as a function of the graph size. The continuous lines corresponds
with our predictions (Tab. \ref{tab1}).}

\label{fig4}
\end{figure}

Our predictions for the cycle abundance are based on centrally connected
cycles, in which a central vertex is connected to all vertices of the
cycle (Fig. \ref{fig1}b). In the following we show that our predictions
capture the scaling of all $h$-cycles as well, not only those that are
centrally connected. For this in Fig. \ref{fig4} we plot the number of
$h=3,4,5$ cycles ({\it i.e.} all cycles as well as those that are
centrally connected) as a function of the graph size for the studied real
and model networks, together with our predictions (continuous line).
First we note that in many cases ($h=3$ and 4) the full cycle density and
the density of the centrally connected cycles overlap. In the few cases
($h=5$) where there are systematic differences between the two densities
the $N$-dependence of the two quantities is the same, indicating that our
calculations correctly predict the scaling of all cycles.

For the co-authorship and Internet graphs $\alpha<1$ and $h_c<3$,
therefore the $h=3,4,5$ cycles are predicted to be in the Type I regime
($h>h_c$). In this case $N_h/N\sim N^{\zeta_h}$, where $\zeta_h = \theta
(h-2) + \delta(1-\alpha)(h-h_c)$. For the language graph $\alpha=1$,
therefore $\zeta_h = \theta (h-2)$. For the deterministic model a direct
count of the $h$-cycles reveals that they are of Type II, {\it i.e.} their
density is independent of $N$ \cite{rozenfeld04}, in agreement with our
predictions for $\alpha\geq1$. These predictions are shown as continuous
lines in Fig. \ref{fig4}, indicating a good agreement with the real
measurements.

Our results offer evidence of a quite complex subgraph dynamics. As the
network grows, the density of the Type II subgraphs remains unchanged,
being independent of the system size. In contrast, the density of the Type
I subgraphs increases in an inhomogeneous fashion. Indeed, each ($n$,$t$)  
subgraph has its own growth exponent $\zeta_{nt}$, which means that their
density increases in a differentiated manner: the density of some Type I
subgraphs will grow faster than the density of the other Type I subgraphs.
Thus, inspecting the system at several time intervals one expects
significant shifts in subgraphs densities. As a group, with increasing
network size the Type I graphs will significantly outnumber the constant
density Type II graphs. Therefore the inspection of the graph density at a
given moment will offer us valuable, but limited information about the
overall local structure of a complex network. However, $P(k)$ and the
$C(k)$ functions allow us to predict with high precision the future shifts
in subgraph densities, indicating that a precise knowledge of the global
network characteristics is needed to fully understand the local structure
of the network at any moment.  These results will eventually force us
reevaluate a number of concepts, ranging from the potential
characterization of complex networks based on their subgraph spectrum to
our understanding of the impact of subgraphs on processes taking place on
complex networks \cite{toroczkai04,petermann04}.

\begin{acknowledgments}
A.V. and A.-L.B. were supported by NSF grant DMS 0441089. 
J.G.O. acknowledges financial support of FCT (Portugal), grant
No. SFRH/BD/14168/2003. We wish to thank J.-P. Eckmann and D. Sergi for 
useful discussion on the subject.
\end{acknowledgments}


\begin{thebibliography}{20}
\expandafter\ifx\csname natexlab\endcsname\relax\def\natexlab#1{#1}\fi
\expandafter\ifx\csname bibnamefont\endcsname\relax
  \def\bibnamefont#1{#1}\fi
\expandafter\ifx\csname bibfnamefont\endcsname\relax
  \def\bibfnamefont#1{#1}\fi
\expandafter\ifx\csname citenamefont\endcsname\relax
  \def\citenamefont#1{#1}\fi
\expandafter\ifx\csname url\endcsname\relax
  \def\url#1{\texttt{#1}}\fi
\expandafter\ifx\csname urlprefix\endcsname\relax\def\urlprefix{URL }\fi
\providecommand{\bibinfo}[2]{#2}
\providecommand{\eprint}[2][]{\url{#2}}

\bibitem[{\citenamefont{Shen-Orr et~al.}(2002)\citenamefont{Shen-Orr, Milo,
  Mangan, and Alon}}]{alon02}
\bibinfo{author}{\bibfnamefont{S.~S.} \bibnamefont{Shen-Orr}},
  \bibinfo{author}{\bibfnamefont{R.}~\bibnamefont{Milo}},
  \bibinfo{author}{\bibfnamefont{S.}~\bibnamefont{Mangan}}, \bibnamefont{and}
  \bibinfo{author}{\bibfnamefont{U.}~\bibnamefont{Alon}},
  \bibinfo{journal}{Nat. Genet} \textbf{\bibinfo{volume}{31}},
  \bibinfo{pages}{64} (\bibinfo{year}{2002});
\bibinfo{author}{\bibfnamefont{R.}~\bibnamefont{Milo}},
  \bibinfo{author}{\bibfnamefont{S.~S.} \bibnamefont{Shen-Orr}},
  \bibinfo{author}{\bibfnamefont{S.}~\bibnamefont{Itzkovitz}},
  \bibnamefont{and} \bibinfo{author}{\bibfnamefont{U.}~\bibnamefont{Alon}},
  \bibinfo{journal}{Science} \textbf{\bibinfo{volume}{298}},
  \bibinfo{pages}{824} (\bibinfo{year}{2002}).

\bibitem[{\citenamefont{Wuchty et~al.}(2003)\citenamefont{Wuchty, Oltvai, and
  Barab\'asi}}]{wuchty03}
\bibinfo{author}{\bibfnamefont{S.}~\bibnamefont{Wuchty}},
  \bibinfo{author}{\bibfnamefont{Z.}~\bibnamefont{Oltvai}}, \bibnamefont{and}
  \bibinfo{author}{\bibfnamefont{A.-L.} \bibnamefont{Barab\'asi}},
  \bibinfo{journal}{Nat. Genet.} \textbf{\bibinfo{volume}{35}},
  \bibinfo{pages}{118} (\bibinfo{year}{2003}).

\bibitem[{\citenamefont{Ulanowicz}(1983)}]{ulanowicz83}
\bibinfo{author}{\bibfnamefont{R.~E.} \bibnamefont{Ulanowicz}},
  \bibinfo{journal}{Math. Bioscienc.} \textbf{\bibinfo{volume}{65}},
  \bibinfo{pages}{219} (\bibinfo{year}{1983}).

\bibitem[{\citenamefont{Bearman}(1997)}]{bearman97}
\bibinfo{author}{\bibfnamefont{P.}~\bibnamefont{Bearman}},
  \bibinfo{journal}{Am. J. Soc.} \textbf{\bibinfo{volume}{102}},
  \bibinfo{pages}{1383} (\bibinfo{year}{1997}).

\bibitem[{\citenamefont{Bernstein}(1999)}]{bernstein00}
\bibinfo{author}{\bibfnamefont{M.}~\bibnamefont{Bernstein}},
  \bibinfo{journal}{ACM Computing Surveys} \textbf{\bibinfo{volume}{31}}
  (\bibinfo{year}{1999}).

\bibitem[{\citenamefont{Marinari and Monasson}()}]{marinari04}
\bibinfo{author}{\bibfnamefont{E.}~\bibnamefont{Marinari}} \bibnamefont{and}
  \bibinfo{author}{\bibfnamefont{R.}~\bibnamefont{Monasson}},
  \bibinfo{note}{arXive:cond-mat/0407253}.

\bibitem[{\citenamefont{V\'azquez et~al.}()\citenamefont{V\'azquez, Dobrin,
  Sergi, Eckmann, Oltvai, and Barab\'asi}}]{vazquez04}
\bibinfo{author}{\bibfnamefont{A.}~\bibnamefont{V\'azquez}},
  \bibinfo{author}{\bibfnamefont{R.}~\bibnamefont{Dobrin}},
  \bibinfo{author}{\bibfnamefont{D.}~\bibnamefont{Sergi}},
  \bibinfo{author}{\bibfnamefont{J.-P.} \bibnamefont{Eckmann}},
  \bibinfo{author}{\bibfnamefont{Z.}~\bibnamefont{Oltvai}}, \bibnamefont{and}
  \bibinfo{author}{\bibfnamefont{A.-L.} \bibnamefont{Barab\'asi}},
  \bibinfo{journal}{PNAS} 
\textbf{\bibinfo{volume}{101}},
  \bibinfo{pages}{17940} (\bibinfo{year}{2004}).


\bibitem[{\citenamefont{Bianconi et~al.}({\natexlab{a}})\citenamefont{Bianconi,
  Caldarelli, and Capocci}}]{bianconi03a}
\bibinfo{author}{\bibfnamefont{G.}~\bibnamefont{Bianconi}},
  \bibinfo{author}{\bibfnamefont{G.}~\bibnamefont{Caldarelli}},
  \bibnamefont{and} \bibinfo{author}{\bibfnamefont{A.}~\bibnamefont{Capocci}},
  \bibinfo{note}{arXiv:cond-mat/0310339}.

\bibitem[{\citenamefont{Bianconi et~al.}({\natexlab{b}})\citenamefont{Bianconi,
  Caldarelli, and Capocci}}]{bianconi04}
\bibinfo{author}{\bibfnamefont{G.}~\bibnamefont{Bianconi}},
  \bibinfo{author}{\bibfnamefont{G.}~\bibnamefont{Caldarelli}},
  \bibnamefont{and} \bibinfo{author}{\bibfnamefont{A.}~\bibnamefont{Capocci}},
  \bibinfo{note}{arXiv:cond-mat/0408349}.

\bibitem[{\citenamefont{Rozenfeld et~al.}()\citenamefont{Rozenfeld, Kirk,
  Bollt, and ben Avraham}}]{rozenfeld04}
\bibinfo{author}{\bibfnamefont{H.~D.} \bibnamefont{Rozenfeld}},
  \bibinfo{author}{\bibfnamefont{J.~E.} \bibnamefont{Kirk}},
  \bibinfo{author}{\bibfnamefont{E.~M.} \bibnamefont{Bollt}}, \bibnamefont{and}
  \bibinfo{author}{\bibfnamefont{D.}~\bibnamefont{ben Avraham}},
  \bibinfo{note}{arXiv:cond-mat/0403536}.

\bibitem{sergi04} D. Sergi, arXive:cond-mat/0412472.

\bibitem[{\citenamefont{Albert and Barab\'asi}(2001)}]{ab01a}
\bibinfo{author}{\bibfnamefont{R.}~\bibnamefont{Albert}} \bibnamefont{and}
  \bibinfo{author}{\bibfnamefont{A.-L.} \bibnamefont{Barab\'asi}},
  \bibinfo{journal}{Rev. Mod. Phys.} \textbf{\bibinfo{volume}{74}},
  \bibinfo{pages}{47} (\bibinfo{year}{2001}).

\bibitem[{\citenamefont{Barab\'asi et~al.}(2002)\citenamefont{Barab\'asi,
  Jeong, Z.N\'eda, Ravasz, Schubert, and Vicsek}}]{bjnr01a}
\bibinfo{author}{\bibfnamefont{A.-L.} \bibnamefont{Barab\'asi}},
  \bibinfo{author}{\bibfnamefont{H.}~\bibnamefont{Jeong}},
  \bibinfo{author}{\bibnamefont{Z.N\'eda}},
  \bibinfo{author}{\bibfnamefont{E.}~\bibnamefont{Ravasz}},
  \bibinfo{author}{\bibfnamefont{A.}~\bibnamefont{Schubert}}, \bibnamefont{and}
  \bibinfo{author}{\bibfnamefont{T.}~\bibnamefont{Vicsek}},
  \bibinfo{journal}{Physica A} \textbf{\bibinfo{volume}{311}},
  \bibinfo{pages}{590} (\bibinfo{year}{2002}).

\bibitem[{\citenamefont{Pastor-Satorras
  et~al.}(2001)\citenamefont{Pastor-Satorras, V\'azquez, and
  Vespignani}}]{pvv01}
\bibinfo{author}{\bibfnamefont{R.}~\bibnamefont{Pastor-Satorras}},
  \bibinfo{author}{\bibfnamefont{A.}~\bibnamefont{V\'azquez}},
  \bibnamefont{and}
  \bibinfo{author}{\bibfnamefont{A.}~\bibnamefont{Vespignani}},
  \bibinfo{journal}{Phys. Rev. Lett.} \textbf{\bibinfo{volume}{87}},
  \bibinfo{pages}{258701} (\bibinfo{year}{2001}).

\bibitem[{\citenamefont{V\'azquez et~al.}(2002)\citenamefont{V\'azquez,
  Pastor-Satorras, and Vespignani}}]{vpv02a}
\bibinfo{author}{\bibfnamefont{A.}~\bibnamefont{V\'azquez}},
  \bibinfo{author}{\bibfnamefont{R.}~\bibnamefont{Pastor-Satorras}},
  \bibnamefont{and}
  \bibinfo{author}{\bibfnamefont{A.}~\bibnamefont{Vespignani}},
  \bibinfo{journal}{Phys. Rev. E} \textbf{\bibinfo{volume}{65}},
  \bibinfo{pages}{066130} (\bibinfo{year}{2002}).

\bibitem[{\citenamefont{Yook and Jeong}()}]{yook}
\bibinfo{author}{\bibfnamefont{S.~H.} \bibnamefont{Yook}} \bibnamefont{and}
  \bibinfo{author}{\bibfnamefont{H.}~\bibnamefont{Jeong}},
  \bibinfo{note}{(unpublished)}.

\bibitem[{\citenamefont{Dorogovtsev et~al.}(2002)\citenamefont{Dorogovtsev,
  Goltsev, and Mendes}}]{dgm02a}
\bibinfo{author}{\bibfnamefont{S.~N.} \bibnamefont{Dorogovtsev}},
  \bibinfo{author}{\bibfnamefont{A.~V.} \bibnamefont{Goltsev}},
  \bibnamefont{and} \bibinfo{author}{\bibfnamefont{J.~F.~F.}
  \bibnamefont{Mendes}}, \bibinfo{journal}{Phys. Rev. E}
  \textbf{\bibinfo{volume}{65}}, \bibinfo{pages}{066122}
  (\bibinfo{year}{2002}).

\bibitem[{\citenamefont{Bianconi and Capocci}(2003)}]{bianconi03}
\bibinfo{author}{\bibfnamefont{G.}~\bibnamefont{Bianconi}} \bibnamefont{and}
  \bibinfo{author}{\bibfnamefont{A.}~\bibnamefont{Capocci}},
  \bibinfo{journal}{Phys. Rev. Lett.} \textbf{\bibinfo{volume}{90}},
  \bibinfo{pages}{78701} (\bibinfo{year}{2003}).

\bibitem[{\citenamefont{Toroczkai et~al.}(2004)\citenamefont{Toroczkai, Kozma,
  Bassler, Hengartner, and Korniss}}]{toroczkai04}
\bibinfo{author}{\bibfnamefont{Z.}~\bibnamefont{Toroczkai}},
  \bibinfo{author}{\bibfnamefont{B.}~\bibnamefont{Kozma}},
  \bibinfo{author}{\bibfnamefont{K.~E.} \bibnamefont{Bassler}},
  \bibinfo{author}{\bibfnamefont{N.}~\bibnamefont{Hengartner}},
  \bibnamefont{and} \bibinfo{author}{\bibfnamefont{G.}~\bibnamefont{Korniss}},
  \bibinfo{journal}{Nature} \textbf{\bibinfo{volume}{428}},
  \bibinfo{pages}{716} (\bibinfo{year}{2004}).

\bibitem[{\citenamefont{Petermann and de~los Rios}(2004)}]{petermann04}
\bibinfo{author}{\bibfnamefont{T.}~\bibnamefont{Petermann}} \bibnamefont{and}
  \bibinfo{author}{\bibfnamefont{P.}~\bibnamefont{de~los Rios}},
  \bibinfo{journal}{Phys. Rev. E} \textbf{\bibinfo{volume}{69}},
  \bibinfo{pages}{066116} (\bibinfo{year}{2004}).

\end{thebibliography}
\end{document}